# Predicting rice blast disease: machine learning versus process based models

David F. Nettleton[1], Dimitrios Katsantonis[2], Argyris Kalaitzidis[2], Natasa Sarafijanovic-Djukic[1], Pau Puigdollers[3] and Roberto Confalonieri[4]

Addresses: [1]IRIS Advanced Engineering, Parc Mediterrani de la Tecnologia, Avda. Carl Friedrich Gauss nº 11, 08860 Castelldefels, Spain, [2]Hellenic Agricultural Organization-DEMETER, Institute of Plant Breeding and Genetic Resources, 65, Georgikis Scholis Av. Zeda Building, Entrance 4, 2nd floor, 57001 Thessaloniki, Greece, [3]GreenPowerMonitor, Av. de Josep Tarradellas, 123-127, 08029 Barcelona – Spain and [4]Università degli Studi di Milano, ESP, Cassandra Lab., Via Celoria, 2, 20133 Milan, Italy.

Emails: David F. Nettleton* - david.nettleton@iris.cat; Dimitrios Katsantonis - dikatsa@cerealinstitute.gr; Argyris Kalaitzidis- kalaitzidis@cerealinstitute.gr; Natasa Sarafijanovic-Djukic - natasa.sdj@iris.cat; Pau Puigdollers - pau.puigdollers@greenpowermonitor.com; Roberto Confalonieri - roberto.confalonieri@unimi.it

* Corresponding author

## Abstract

**Background:** Rice is the second most important cereal crop worldwide, and the first in terms of number of people who depend on it as a major staple food. Rice blast disease is the most important biotic constraint of rice cultivation causing each year millions of dollars of losses. Despite the efforts for breeding new resistant varieties, agricultural practices and chemical control are still the most important methods for disease management. Thus, rice blast forecasting is a primary tool to support rice growers in controlling the disease. In this study, we compared four models for predicting rice blast disease, two operational process-based models (Yoshino and WARM) and two approaches based on machine learning algorithms (M5Rules and RNN), the former inducing a rule-based model and the latter building a neural network. In situ telemetry is important to obtain quality in-field data for predictive models and this was a key aspect of the RICE-GUARD project on which this study is based. According to the authors, this is the first time process-based and machine learning modelling approaches for supporting plant disease management are compared.

**Results:** Results clearly showed that the models succeeded in providing a warning of rice blast onset and presence, thus representing suitable solutions for preventive remedial actions targeting the mitigation of yield losses and the reduction of fungicide use. Statistical metrics were derived to quantify the ability of the models to anticipate rice blast onset and detect its presence. For this, the real blast severity was used as the output for training and testing the models. The Area Under Curve (AUC) was used as a metric to quantify the "early warning" success during the early period of the crop growing season. The M5Rules, RNN and Yoshino models all gave significant "signals" during the "early warning" period (15th/20th June to 7th July), with average AUC values of 214, 181 and 163, respectively. WARM also gave some signals during this period but of a spiky form so the AUC value of 39 was relatively lower for this reason. The best average values of %MAE, R and $R^2$, for the machine learning models were 0.78, 0.61 and 0.72, respectively and for the process models the corresponding values were 0.65, 0.46 and 0.36. This has relevant implications for the operational use of the models, since most of the available studies limited to analyse the relationship between model outputs and the incidence of rice blast.

Results also showed that machine learning methods approximated the performances of two process-based models used for years in operational contexts.

**Conclusions:** Process-based and data-driven models can be used to provide early warnings to anticipate rice blast and detect its presence, thus supporting fungicide applications. Data-driven models derived from machine learning methods are a viable alternative to process-based approaches and – in cases when training datasets are available – offer a potentially greater adaptability to new contexts.

## Background

Rice (*Oryza sativa* L.), after wheat, is a major staple crop for more than half of the world's population [1], with more than 3.5 billion people depending on rice for more than 20% of their calories demand. This includes 70% of the world's 1.3 billion poorest who live in Asia, where rice is the predominant crop. In Europe, it has been cultivated for centuries mainly throughout the Mediterranean countries: Italy, Spain, Greece, Portugal and France [2]. The most critical constraint limiting rice productions worldwide is blast disease, caused by *Pyricularia oryzae* Cavara [3]. The rice blast fungus is capable of infecting plants at different stages: it appears early on as white/grey and brownish leaf lesions, followed later by nodal rot and neck blast, which can cause necrosis and often breakage of the panicle (compound raceme or branched cluster of flowers) [4]. In Figure 1, four different grades of leaf lesions are shown. At present, the fungus can be found in over 85 countries worldwide [5], being the most important rice disease in China, Japan and USA, where it can cause severe yield losses [6,7,8]. It is estimated that a moderate infection in the field is enough to cause a 50% reduction in yield. Devi and Sharma [9] estimated that the fungus is capable of destroying annually enough rice to feed 60 million people.

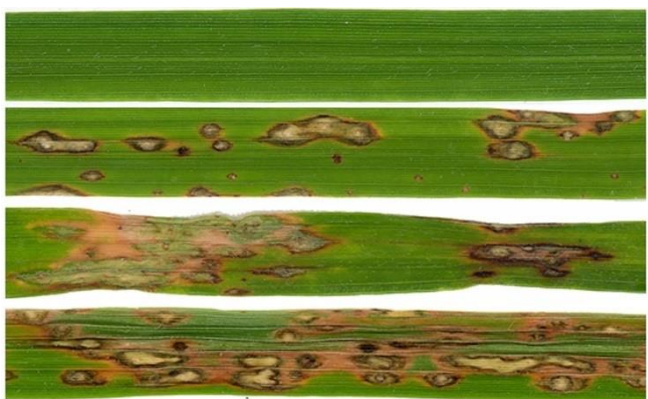

**Figure 1.** Rice Blast - different grades of leaf lesion (source: http://journals.plos.org/plosone/article?id=10.1371/journal.pone.0026260 )

The management of blast disease has been extensively investigated by many researchers in several countries [10,11,12,13,14,15,16,17,18,19,20,21]. Despite all the efforts, rice blast has never been fully eliminated from a region in which rice is grown - a single change in practices or the way in which resistant genes are deployed can result in a return to disease presence even after many years of successful management [22]. Thus, fungicide applications still remain the most effective method for controlling rice blast, despite raising doubts on the environmental impact of chemicals and on their role in inducing fungicide resistance within the pathogen populations [23].

Among the methods to manage and control a major disease like rice blast, a key role is played by forecasting systems. Disease forecasts can indeed assist farmers and other end-users to make strategic decisions concerning the number and timing of fungicide applications, define fertilization practices by avoiding luxury consumption (in turn increasing plant susceptibility), and even to predict yields [24]. However, in biological terms, forecasting systems are based on assumptions concerning the pathogen's interactions with the host and the environment, which are widely known as the "disease triangle", whose three sides are: a) "favorable conditions", b) "virulent pathogen" and c) "susceptible host" [25]. The availability of robust and reliable early-warning systems would allow preventing the explosive nature of the disease through the timely application of control measures [26]. This would turn into the reduction of both yield losses and fungicide applications, thus minimizing the environmental footprint of rice cultivation.

Katsantonis et al. [4] conducted a comprehensive review of 52 rice blast prediction models developed and used worldwide, which highlighted the approaches from Yoshino [27] and from the WARM rice model [28,29] as characterized by a good potential for operational applications. Yoshino represents one of the earliest attempts in rice blast prediction modelling and it has been widely incorporated in many operational alert systems. WARM is the result of more recent researches, and it is part of the EU service "Monitoring Agricultural ResourceS" (MARS) and of operational early warning systems used in Italy [24]. This led to consider the Yoshino and WARM approaches as benchmarking systems for the current study.

One the other hand, the literature on rice blast prediction using machine learning and statistical techniques is relatively new – the first and most referenced research is likely the one provided by Kaundal et al. [26]. In this study, one statistical and two machine learning techniques – i.e., multiple regression, neural network and support vector machine – were applied to predict rice blast in different sites and seasons. Bregaglio et al. [29] used data derived from laser-induced chlorophyll fluorescence to predict rice blast. They first applied principal components analysis (PCA) to reduce the dimensionality of the spectral information, and then derived statistical models using discriminant analysis (DA), multiple logistic regression analysis (MLRA), and multilayer perceptron (MLP) techniques. They reported an average prediction accuracy of 91.7% using PCA-MLP. Kim et al. [30] predicted rice blast using long short-term memory (LSTM) recurrent neural networks with an accuracy ranging between 40% and 79% across different sites. Malicdem and Fernandez [31] used associative neural network (ANN) and support vector machine (SVM) binary classifiers for predicting occurrence of rice blast. They pre-processed data using PCA to determine the most important weather information. In this study, best performances were obtained with SVM, with mean squared error (MSE) and $R^2$ being 0.23 and 0.77, respectively, for SVM, and 0.46 and 0.47 for ANN.

The aim of the current study was to compare the process-based models Yoshino and WARM with alternative approaches, in turn based on two different machine learning algorithms: M5Rules and RNN.

According to the authors' knowledge, this is the first time process-based models and machine learning approaches are compared using the same dataset. Moreover, besides standard metrics (R, $R^2$ and %MAE) to quantify the agreement between model outputs and incidence of rice blast, we also used the AUC (Area Under Curve) metric to evaluate the models "early warning" success at the start of the rice blast appearance period. This is particularly important in light of the use of rice blast prediction approaches to support fungicide application in operational contexts.

In conclusion to the background section, we would like to state several novel aspects of our work with respect to the state of the art (other novel aspects are mentioned at the end of the Discussion section): (i) the use of a rule induction model (almost all other published research uses 'black box' modelling such as SVM and neural networks) as a machine learning technique. Rule induction provides human readable rules which can be interpreted to give insights into the behaviour and inter-relations between rice blast indicators; (ii) comparison of process models (Yoshino and WARM) with machine learning models (built with M5Rules and RNN) whereas the state of the art (see Background section) compares only process models or only ML built models; (iii) the RICE-GUARD EU project used in-situ state-of-the-art data capture metrology equipment to obtain the datasets used in the present study. This represents an improvement on the data typically available in real scenarios (public meteorological reports), which are often less reliable or more regional in nature.

*The RICE-GUARD Project*
RICE-GUARD [32] is an EU FP7 project aimed at capturing in-field telemetry data to improve the predictive capability of the Yoshino model [27, 14] for rice blast while comparing it with the recent WARM approach [28, 29]. In particular, RICE-GUARD developed a low-cost, in-field wireless sensor network (WSN) to increase the representativeness of the weather data used to feed rice blast forecasting systems. The RICE-GUARD WSN is largely based on advances in the Internet of Things (IoT) technology, which allowed the implementation of wireless networks and radiofrequency communications to collect real-time, spatially distributed weather data (Figure 2). Indeed, although weather data is the main driver of blast models, its reliability is often threatened by the spatial distribution of weather stations, which are often placed outside rice cultivation areas. The resulting uncertainty that often characterizes existing systems for blast alert leads to a lack of confidence in advisory bulletins and to an overuse of fungicides, resulting in sizable economic and environmental costs.

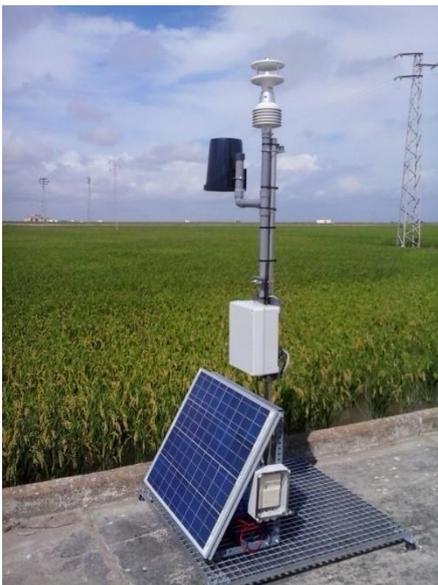

**Figure 2.** Data capture RICE-GUARD station located outside the paddy field, for gathering and transmitting in real time readings from the in-field sensors

**Results**
*Building and training machine learning models*

*(i) M5Rules Rule Induction*
In order to build a dataset for predictive modelling with the Rice Blast Severity index as output, we used the following inputs to the M5Rules algorithm: daily maximums and minimums for air temperature, relative humidity and leaf wetness, together with moving averages for the previous 1, 3 and 7 days of the daily maximums and minimums for air temperature, relative humidity and leaf wetness. The output is a numerical value between 1 and 6 which indicates the Rice Blast

Severity index. Figure 3a shows the complete rule set for the 3x1 M5Rules data model (trained on the k15+s16+p15 datasets) and their tree representation is shown in Figure 3b. With reference to Figure 3a, Rule 3, it can be seen that 18 cases were predicted with 10% training error. Rule 3 uses four moving averages in the "IF" part of the rule (leafwet7, 7 day moving average for leaf wetness; relhum7, 7 day moving average for relative humidity; relhum3, 3 day moving average for relative humidity; relhum1, 1 day moving average for relative humidity) to predict the Rice Blast severity index. In the "THEN" part of the rule the output value is produced for the blast severity, which has 12 components: 0.0008 x temp + 0.001 x leafwet + 0.029 x relhum1, and so on. Overall, it can be seen that the rule model has 7 rules, and in the "IF" part of the rules the most frequently used attributes are relhum7 (7 day moving average of relative humidity) which is used 6 times, and leafwet7 (7 day moving average of leaf wetness) which is used 5 times. It can be seen that although the raw values for the temperature, relative humidity and leaf wetness were included as inputs, they were never used by the model (which uses an "information gain" calculation to choose which attributes to include) and 7 and 3 day moving averages of these values were mainly used. The statistics using 10 fold cross validation (Weka, bottom right of Figure 3a) gave a correlation of 0.9442 with the 264 training instances.

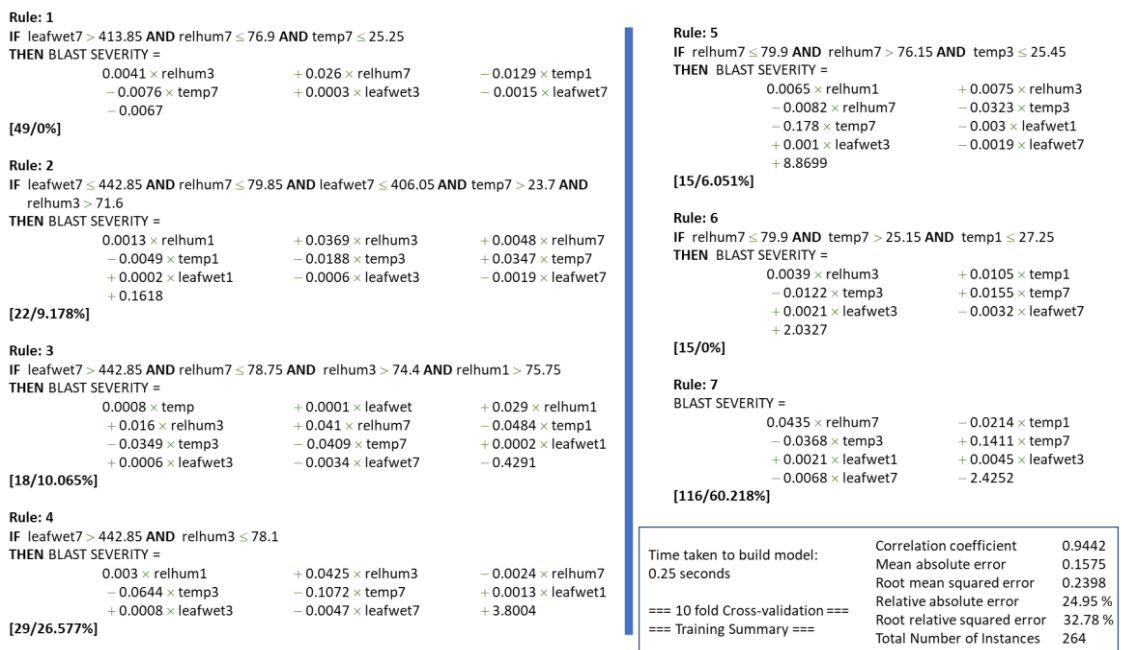

**Figure 3a.** Complete rule set of M5Rules model trained on the k2015+s2016+p2015 datasets.

```
leafwet7 ≤ 413.85 :                                    leafwet7 > 413.85 :
|  temp7 ≤ 23.7 :                                      |  relhum7 ≤ 76.9 :
|  |  relhum7 ≤ 73.5 :                                 |  |  temp7 ≤ 25.25 : LM14 (49/0%)
|  |  |  leafwet7 ≤ 402.15 :                           |  |  temp7 > 25.25 :
|  |  |  |  leafwet7 ≤ 385.35 : LM1 (7/2.868%)         |  |  |  temp7 ≤ 26.2 :
|  |  |  |  leafwet7 > 385.35 :                        |  |  |  |  relhum1 ≤ 73.85 : LM15 (10/10.257%)
|  |  |  |  |  leafwet7 ≤ 390.35 : LM2 (8/4.479%)      |  |  |  |  relhum1 > 73.85 : LM16 (16/4.729%)
|  |  |  |  |  leafwet7 > 390.35 :                     |  |  |  temp7 > 26.2 :
|  |  |  |  |  |  temp7 ≤ 22.2 : LM3 (6/3.613%)        |  |  |  |  temp3 ≤ 26.4 : LM17 (7/2.06%)
|  |  |  |  |  |  temp7 > 22.2 : LM4 (5/8.932%)        |  |  |  |  temp3 > 26.4 : LM18 (5/0%)
|  |  |  leafwet7 > 402.15 :                           |  relhum7 > 76.9 :
|  |  |  |  leafwet3 ≤ 404.15 :                        |  |  relhum7 ≤ 79.9 :
|  |  |  |  |  temp7 ≤ 22.3 : LM5 (4/0%)               |  |  |  temp7 ≤ 25.55 :
|  |  |  |  |  temp7 > 22.3 : LM6 (11/4.35%)           |  |  |  |  leafwet7 ≤ 448.15 : LM19 (8/7.504%)
|  |  |  |  leafwet3 > 404.15 :                        |  |  |  |  leafwet7 > 448.15 : LM20 (9/0%)
|  |  |  |  |  relhum3 ≤ 77.65 : LM7 (7/13.769%)       |  |  |  temp7 > 25.55 : LM21 (22/0%)
|  |  |  |  |  relhum3 > 77.65 : LM8 (3/0.004%)        |  |  relhum7 > 79.9 :
|  |  relhum7 > 73.5 : LM9 (16/3.786%)                 |  |  |  temp3 ≤ 24.9 :
|  temp7 > 23.7 :                                      |  |  |  |  temp3 ≤ 24.3 : LM22 (7/0%)
|  |  relhum3 ≤ 70.15 : LM10 (9/9.348%)                |  |  |  |  temp3 > 24.3 :
|  |  relhum3 > 70.15 :                                |  |  |  |  |  relhum7 ≤ 80.3 : LM23 (5/4.391%)
|  |  |  leafwet1 ≤ 367.5 : LM11 (10/5.86%)            |  |  |  |  |  relhum7 > 80.3 :
|  |  |  leafwet1 > 367.5 :                            |  |  |  |  |  |  relhum3 ≤ 80.8 : LM24 (4/0.655%)
|  |  |  |  leafwet1 ≤ 422.5 : LM12 (13/7.376%)        |  |  |  |  |  |  relhum3 > 80.8 : LM25 (4/0.006%)
|  |  |  |  leafwet1 > 422.5 : LM13 (5/0%)             |  |  |  temp3 > 24.9 : LM26 (14/0%)
```

**Figure 3b.** Tree representation of rule model shown in Figure 3a.

*(ii) Recurrent Neural Networks*

Input data to LSTM RNNs were derived as the daily maximums and minimums for air temperature, relative humidity and leaf wetness. Output is the rice blast severity index (scale of 1 to 6). To implement LSTM RNNs, the Keras library for deep learning [33] was used with TensorFlow as the 'back-end'. Different configurations of LSTM RNNs were explored, such as varying the number of hidden layers for the RNNs and varying the number of LSTM cells in the layers. Moreover, different time windows were tested for the input variables used in the samples given to the RNN. Finally, the simplest configuration was chosen with the smallest time window that gave the minimum required accuracy threshold, in order to avoid over-fitting. This was an RNN with one hidden layer of 10 LSTM cells and a time window of 10 time steps (corresponding to approximately 5 days).

***Running process based and machine learning models***

As our objective was to detect the presence of Rice Blast and obtain an early warning signal for an increase in blast severity, a graphic representation was used which nh clearly shows the model outputs and their degree of correspondence with the real Rice Blast severity index. Figures 5 and 6 illustrate the results for Yoshino, WARM, M5Rules and LSTM RNN models, respectively. In each figure, the real blast severity value is shown, together with the output from the corresponding models and related trend lines, which were then used to calculate MAE and $R^2$ as shown in Tables 1 to 3. In case of considering the trend lines as continuous probability distributions, a probabilistic interpretation can be used as a good approximation for evaluating the presence of Rice Blast. In Figure 5 it can be seen that the Yoshino model triggers before the blast severity starts to increase as well as during the increase and higher risk period. In this section we compare Yoshino (adapted in RICE-GUARD), WARM, M5Rules and LSTM RNN by using four metrics: (i) R (correlation) (ii) R squared (iii) mean absolute error (%MAE); (iv) Area Under Curve (AUC) for period before blast severity starts to rise. For this comparison, Figures 5 and 6 are based on the 3x1 train/test combinations with Kalochori 2016 and Seville 2016 as test datasets, respectively. The AUC is a measure of "early warning" alert during the especially important time period while the rice crop is still young and most susceptible to damage. As a consequence of the alert, preventive spraying actions could be initiated against the rice blast. This is depicted graphically in Figure 4.

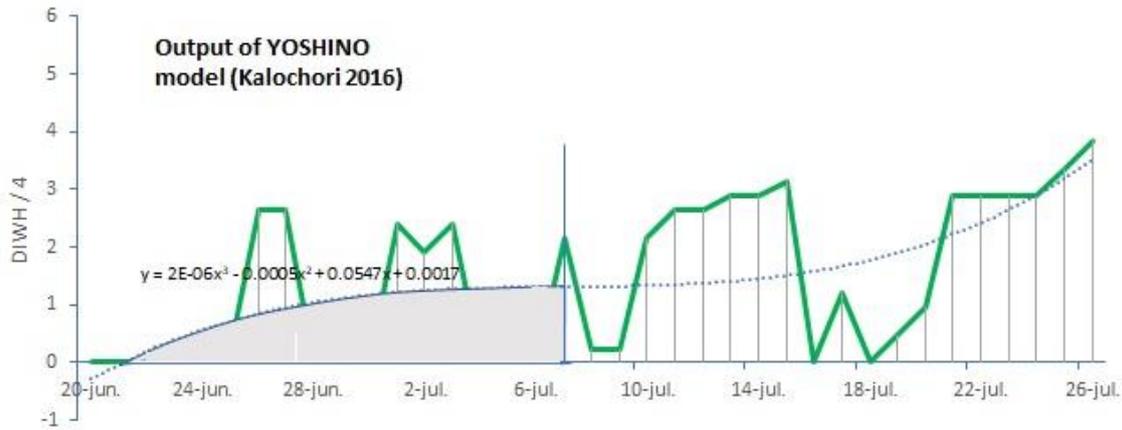

**Figure 4**. AUC (Area Under Curve) used as an "early warning" metric evaluator for the period 20th June to 7th July for the Kalochori 2016 dataset and the Yoshino model output.

With reference to Figure 4, the AUC (area highlighted in red) is measure to quantify the grade to which a model is successful for warning farmers in a timely manner that a rice blast outbreak is imminent, so that they can apply the necessary countermeasures. Figure 4 is a segment of Figure 5 (b), which plots the Yoshino model output. The critical early period is 20th June to 7th July, at which point the blast severity starts to rise (se Figure 5(a)). Hence the AUC serves as a quantification of the "activation level" of the model output during this period.
$AUC$ is calculated as follows:
$$AUC = \int_a^b (2E - 06x^3 - 0.0005x^2 + 0.0547x + 0.0017).dx$$
where a and b is the x-axis range (20th June to 7th July, translated into a numerical sequential index $x = 1..18$) to be evaluated and $2E - 06x^3 - 0.0005x^2 + 0.0547x + 0.0017$ is the equation of the curve whose area is to be calculated. In Table 5, a quantitative evaluation calculated from these metrics is included, derived from the data shown in Figures 5 and 6. This is interpreted later in this section. For calculation of the early warning time period we took from the start date of the spraying season to the date the blast severity started to rise. This was based on the empirical study of the Kalochori and Seville sites, that approximately reflects useful time of the spraying (of course this can be customized according to specific and/or local agronomic information).

Tables 1 to 3 show the results for the M5Rules rule induction algorithm, for four different datasets (two different time-periods and three different locations).
Table 1 shows the results for models built from different combinations of unique datasets for train and test. For example, in row 1, k2016 (Kalochori 2016) is the training dataset and k2015 (Kalochori 2015) is the test dataset. Unique training datasets, as expected, were found to be the most difficult to build models from, given the limited generalization from one location to another and from one year to another. With reference to Table 1, the average values of the r, MAE and $r^2$, for the 1x1 M5Rules models were 0.49, 0.59 and 0.27, respectively and for the 1x1 RNN models, 0.54, 0.62 and 0.32, respectively.

Table 2 shows the results for models built from different combinations of three datasets for training and one for testing. For example, in row 1, k2015 (Kalochori 2015), s2016 (Seville 2016) and p2015 (Portugal 2015) are used as training datasets and k2016 (Kalochori 2016) as the test dataset. The 3x1 combinations, as expected, were found to give the best models, given the greater generalization capability of the training data from different locations and years. With reference to Table 2, the average values of the r, MAE and $r^2$, for the 3x1 M5Rules models were 0.60, 0.63 and 0.40, respectively and for the 3x1 RNN models, 0.70, 0.75 and 0.51, respectively. Table 3 shows the results for models built from different combinations of two datasets for training datasets and one for testing. For example, in row 1, k2015 (Kalochori 2015) together with k2016 (Kalochori 2016) are used as training datasets and s2016 (Seville 2016) is the test dataset. The 2x1 combinations, as expected, were found to give a model quality in between the unique dataset models of Table 1 and 3x1 models of Table 2, for the same reasons as previously given. With reference to Table 3, the average values of the r, MAE and $r^2$, for the 2x1 M5Rules models were 0.42, 0.73 and 0.21, respectively and for the 2x1 RNN models, 0.57, 0.67 and 0.37. The results also reflect some of the data quality issues due to difficulties experienced during the in-field data capture: different devices, sensor failures. Also, the differences between different locations in Europe which although similar, each having its own "micro-climate" in terms of temperature, humidity and leaf wetness and their relation to the incidence/severity of rice blast.

**Table 1** Comparison of performance of M5Rules and RNN models on different individual dataset combinations.

| | | Train | | Test | | | | | |
|---|---|---|---|---|---|---|---|---|---|
| Training data* | Test data | r (a) | r (b) | r (a) | r (b) | MAE** (a) | MAE** (b) | $r^2$ (a) | $r^2$ (b) |
| k2016 | k2015 | 0.91 | 0.84 | 0.62 | 0.81 | 0.55 | 0.54 | 0.38 | 0.66 |
| k2016 | s2016 | | | 0.78 | 0.67 | 0.41 | 0.85 | 0.61 | 0.44 |
| k2016 | p2015 | | | 0.37 | 0.43 | 0.62 | 0.25 | 0.13 | 0.18 |
| k2015 | k2016 | 0.95 | 0.92 | 0.68 | 0.60 | 0.53 | 0.68 | 0.46 | 0.36 |
| k2015 | s2016 | | | 0.64 | 0.70 | 0.52 | 0.61 | 0.40 | 0.49 |
| k2015 | p2015 | | | 0.31 | 0.29 | 0.69 | 0.82 | 0.10 | 0.08 |
| s2016 | k2016 | 0.98 | 0.87 | 0.40 | 0.57 | 0.65 | 0.59 | 0.16 | 0.32 |
| s2016 | k2015 | | | 0.38 | 0.66 | 0.61 | 0.59 | 0.14 | 0.44 |
| s2016 | p2015 | | | 0.24 | 0.21 | 0.68 | 0.75 | 0.06 | 0.04 |
| p2015 | k2016 | 0.80 | 0.75 | 0.67 | 0.65 | 0.57 | 0.63 | 0.45 | 0.42 |
| p2015 | k2015 | | | 0.44 | 0.49 | 0.46 | 0.38 | 0.19 | 0.24 |
| p2015 | s2016 | | | 0.35 | 0.45 | 0.76 | 0.80 | 0.12 | 0.20 |
| **Average values** | | | | **0.49** | **0.54** | **0.59** | **0.62** | **0.27** | **0.32** |

*k=Kalochori, s=Seville, p=Portugal; **mean absolute error, (a)=M5Rules, (b)=RNN

**Table 2** Comparison of performance of M5Rules and RNN models on different 3 × 1 dataset combinations.

| | | Train | | Test | | | | | |
|---|---|---|---|---|---|---|---|---|---|
| Training data* | Test data | r (a) | r (b) | r (a) | r (b) | MAE** (a) | MAE** (b) | $r^2$ (a) | $r^2$ (b) |
| k2015+s2016+p2015 | k2016 | 0.94 | 0.96 | 0.76 | 0.81 | 0.52 | 0.49 | 0.58 | 0.66 |
| k2016+s2016+p2015 | k2015 | 0.89 | 0.91 | 0.29 | 0.72 | 0.49 | 0.58 | 0.09 | 0.52 |
| k2016+k2015+p2015 | s2016 | 0.88 | 0.86 | 0.77 | 0.75 | 0.88 | 0.95 | 0.60 | 0.56 |
| k2016+k2015+s2016 | p2015 | 0.91 | 0.87 | 0.57 | 0.53 | 0.62 | 0.99 | 0.32 | 0.28 |
| **Average Values** | | | | **0.60** | **0.70** | **0.63** | **0.75** | **0.40** | **0.51** |

*k=Kalochori, s=Seville, p=Portugal; **mean absolute error, (a)=M5Rules, (b)=RNN

**Table 3** Comparison of performance of M5Rules and RNN models on different 2 × 1 dataset combinations.

| | | Train | | Test | | | | | |
|---|---|---|---|---|---|---|---|---|---|
| Training data* | Test data | r (a) | r (b) | r (a) | r (b) | MAE** (a) | MAE** (b) | $r^2$ (a) | $r^2$ (b) |
| k2015+k2016 | s2016 | 0.94 | 0.94 | 0.56 | 0.72 | 0.78 | 0.63 | 0.31 | 0.52 |
| k2015+k2016 | p2015 | | | 0.21 | 0.31 | 0.60 | 0.88 | 0.04 | 0.10 |
| k2015+p2015 | k2016 | 0.87 | 0.85 | 0.69 | 0.68 | 0.78 | 0.53 | 0.48 | 0.46 |
| k2015+p2015 | s2016 | | | 0.67 | 0.57 | 0.91 | 0.71 | 0.45 | 0.32 |
| k2015+s2016 | k2016 | 0.92 | 0.95 | 0.46 | 0.71 | 0.39 | 0.51 | 0.21 | 0.66 |
| k2015+s2016 | p2015 | | | 0.22 | 0.25 | 1.04 | 1.10 | 0.05 | 0.06 |
| k2016+p2015 | k2015 | 0.92 | 0.94 | 0.61 | 0.69 | 0.16 | 0.48 | 0.37 | 0.47 |
| k2016+p2015 | s2016 | | | 0.52 | 0.58 | 0.98 | 0.69 | 0.27 | 0.33 |
| k2016+s2016 | k2015 | 0.89 | 0.92 | 0.36 | 0.73 | 0.40 | 0.52 | 0.13 | 0.62 |
| k2016+s2016 | p2015 | | | 0.17 | 0.35 | 1.20 | 0.78 | 0.03 | 0.12 |
| s2016+p2015 | k2015 | 0.91 | 0.94 | 0.33 | 0.63 | 0.80 | 0.63 | 0.11 | 0.40 |
| s2016+p2015 | k2016 | | | 0.19 | 0.57 | 0.72 | 0.62 | 0.04 | 0.32 |
| | Average | | | **0.42** | **0.57** | **0.73** | **0.67** | **0.21** | **0.37** |

*k=Kalochori, s=Seville, p=Portugal; **mean absolute error, (a)=M5Rules, (b)=RNN

**Table 4** Process models vs ML models (3x1 combinations)

| | M5RULES | | | RNN | | | YOSHINO | | | WARM | | |
|---|---|---|---|---|---|---|---|---|---|---|---|---|
| Dataset | R | $R^2$ | %MAE* | R | $R^2$ | %MAE* | R | $R^2$ | %MAE* | R | $R^2$ | %MAE* |
| K2016 | 0.76 | 0.58 | 0.52 | 0.81 | 0.66 | 0.49 | 0.84 | 0.71 | 0.23 | -0.72 | 0.53 | 0.77 |
| S2016 | 0.77 | 0.60 | 0.88 | 0.75 | 0.56 | 0.95 | 0.47 | 0.22 | 0.50 | 0.00 | 0.00 | 0.00 |
| K2015 | 0.29 | 0.09 | 0.49 | 0.72 | 0.52 | 0.58 | 0.34 | 0.12 | 0.75 | 0.78 | 0.61 | 0.76 |
| P2015 | 0.57 | 0.32 | 0.62 | 0.53 | 0.28 | 0.99 | 0.00 | 0.00 | 0.00 | 0.69 | 0.48 | 0.92 |
| Avg. | 0.59 | 0.39 | 0.63 | 0.70 | 0.50 | 0.75 | 0.41 | 0.26 | 0.37 | 0.19 | 0.40 | 0.61 |

For the results shown in Table 4, it can be seen that the ML models are competitive with the process models in terms of the fit (R, R2 and %MAE) to the real blast severity. RNN has the best average R value (0.70), followed by M5Rules (0.59), Yoshino (0.41) and WARM (0.19). The process model average precisions were affected by the zero valued results for p2015 (Yoshino) and s2016 (WARM). WARM gave relatively good results for k2015 and p2015 but Yoshino gave better performance for three of the four datasets. The zero values indicated that Yoshino and WARM were unable to give an output for these datasets and time periods. One possible cause is the lack of rainfall data for these cases, Also, there were differences in the two LW sensors (GR, Hobo and PT, Pentagon) and it was necessary to define different thresholds for each. So, from the results in Table 4 it can be seen that the ML models were more robust to changes in data or data availability, whereas the process models were more sensitive to changes in the available data (i.e. in the cases of S2016 and P2015). Note that in Table 4, the M5Rules and RNN results have been chosen from Tables 3 as the ones with the best R values for the corresponding test datasets. Furthermore, in the case of WARM, due to the spiky nature of the model output, the real utility

(for all models) in the field as support to the farmers is actually greater than the R values suggest. This is because the spike of the output acts as a trigger/alert, and is especially effective when it occurs before the blast severity starts to increase. This is actually the case for all four models, as is described in relation to Figures 5 and 6.

With reference to Table 5, in order to evaluate the models for their utility as tools for providing farmers with early warnings the "early warning success" metric was applied as the AUC (Area Under Curve) as explained previously. That is, the AUC was calculated from 20$^{th}$ June to 7$^{th}$ July for the Kalochori 2016 dataset (Figs 5), from 15$^{th}$ June to 7$^{th}$ July for the Seville 2016 dataset (Figs 6), from 12$^{th}$ to 14$^{th}$ August for the Kalochori 2015 dataset and from 5$^{th}$ to 9$^{th}$ August for the Portugal 2015 dataset. It appears that NN, M5Rules and Yoshino all gave significant "signals" during the "early warning" period. WARM also gave some signals during this period but of a spiky form so the AUC value is significantly lower for this reason.

**Table 5** Comparison of performance of models in terms of "early warning success" with data sets from Kalochori 2016 and Seville 2016

| | Early warning Success | | | | |
|---|---|---|---|---|---|
| **Model** | **Kalochori 2016** | **Seville 2016** | **Kalochori 2015** | **Portugal 2015** | **Average** |
| Yoshino | 151 | 176 | 26.16 | 0** | 88.29 |
| WARM | 39 | 0** | -1.38 | -0.62 | 9.25 |
| M5Rules | 213 | 216 | 6.2 | 12.27 | 111.87 |
| LSTM NN | 169 | 193 | -0.86 | 13.67 | 181 |

*AUC, Area Under Curve, **Gave zero values for whole time range*

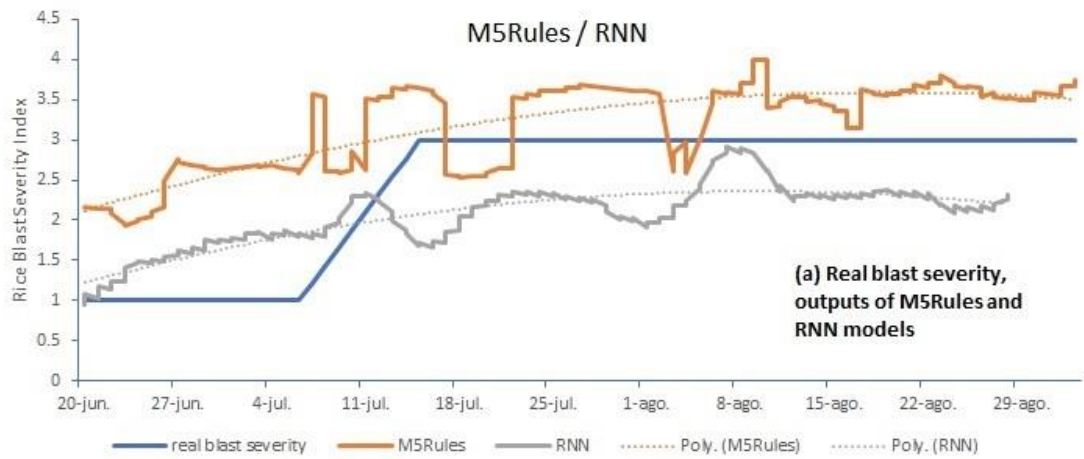

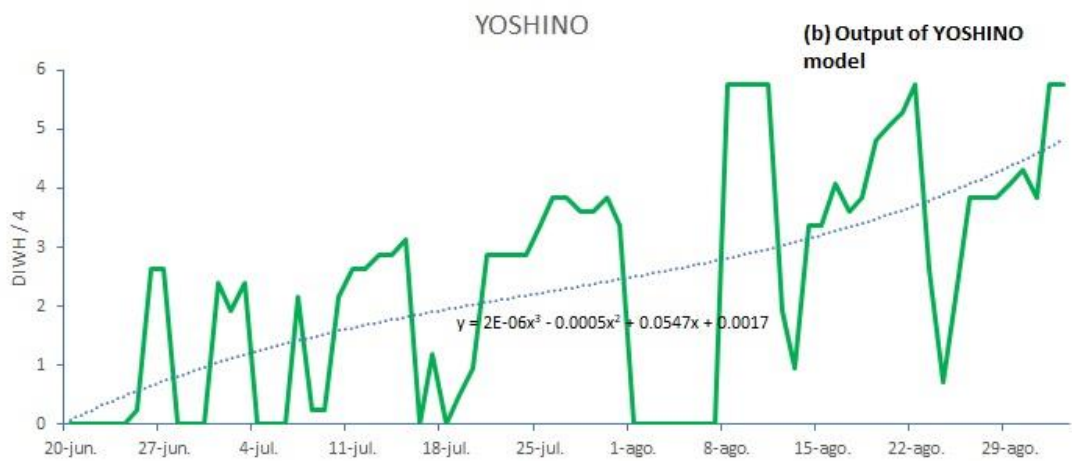

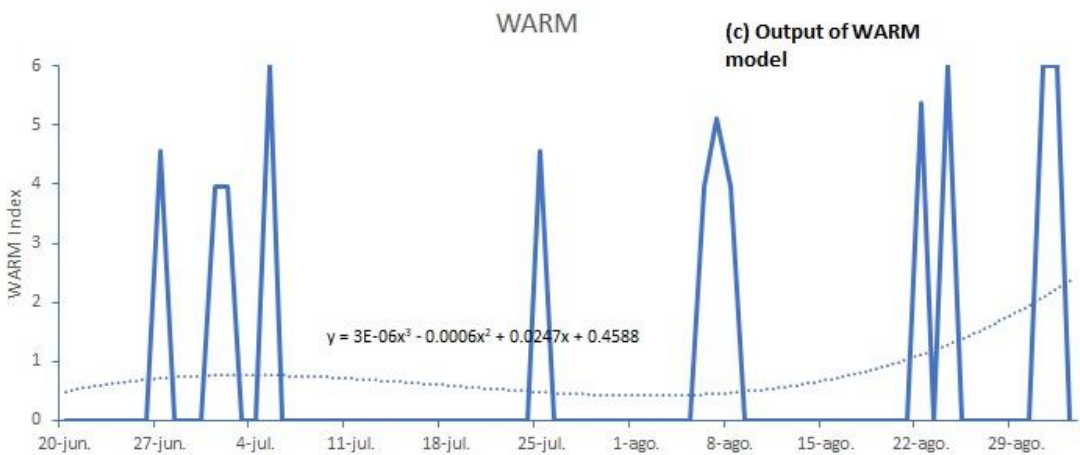

**Figure 5.** Real blast severity vs predicted: Train k2015+s2016+p2015, Test k2016

With reference to Fig. 6, it can be seen that the predicted blast severity value again starts to rise before the real onset of the rice blast, going from an average of 2 to 3, and then stays at a level 3 as the real blast severity also reaches 3.

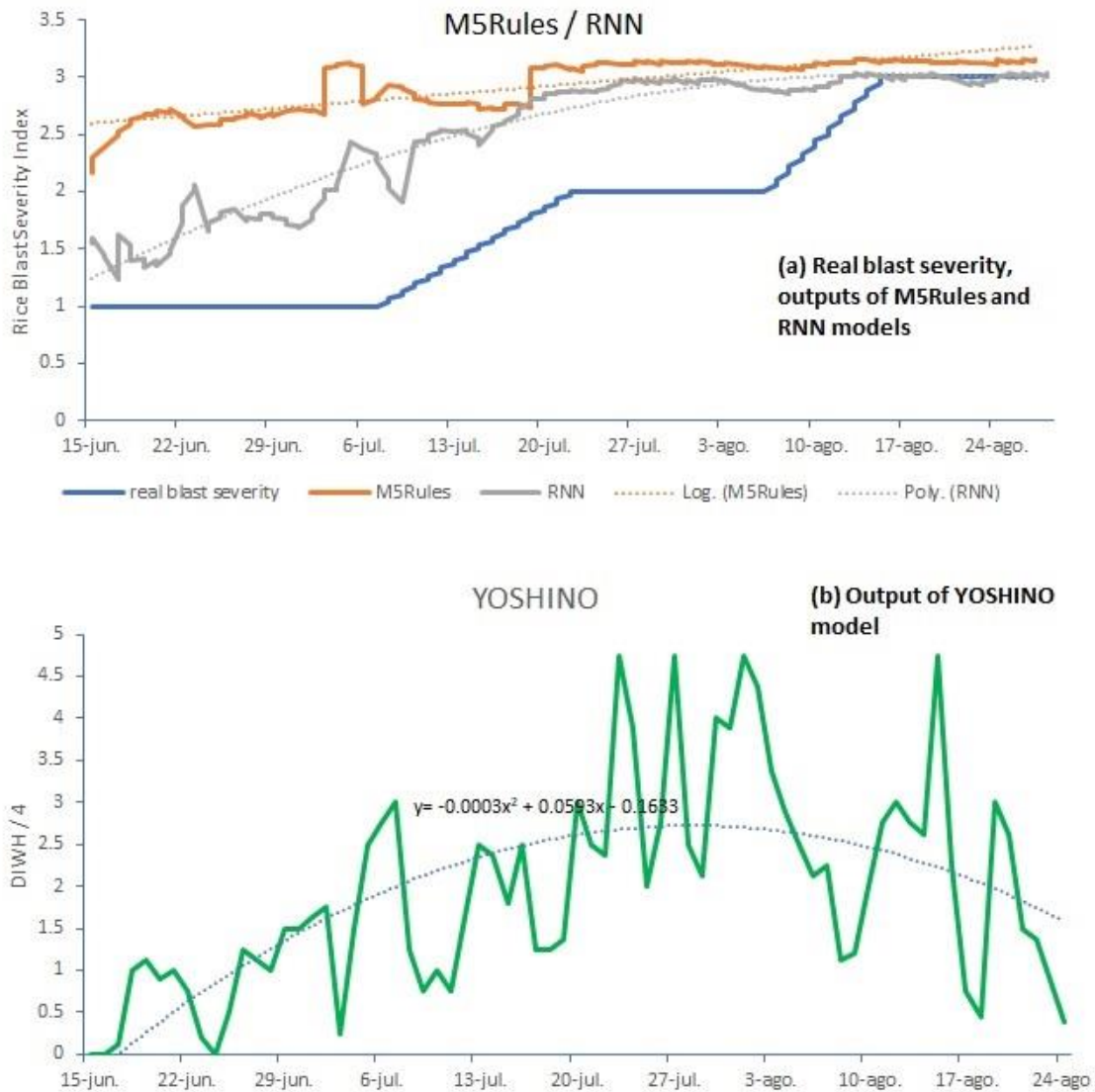

**Figure 6**. Real blast severity vs predicted: Train k2016+k2015+p2015, Test s2016

**Discussion**

All the approaches under evaluation, i.e., Yoshino, WARM and the two machine learning approaches (M5Rules and LSTM RNNs), succeeded in providing warnings on onset and presence of rice blast early enough to allow farmers taking necessary measures. This can be seen from Figs. 5 and 6, where the output of all models triggers before the blast severity actually starts to rise around 7[th] July for the Kalochori and Seville sites in 2016. So, given the Yoshino and WARM models are already available and effective, what would be the advantage of using the machine learning approaches? Yoshino and WARM models are based on a fixed model structure and contain parameters that need to be defined by experts using detailed datasets of observations, given their values could vary across geographic areas and climates. Indeed, within the RICE-GUARD project, a different temperature threshold was used in the Yoshino model to adapt it to the conditions experienced by the pathogen in the Mediterranean climate, because Yoshino was originally developed and tested in Asia. The use of data-driven machine learning algorithms makes it easier to customize the resulting rice blast models to specific areas and climates.

Concerning machine learning approaches, our results are similar to those achieved by Kaundal et al. [26], who used machine learning and statistical techniques to predict rice blast. For cross-locations training/test they obtained $R^2$ ranging between 0.01 and 0.98 (their best approach was the Support Vector Machine, SVM), and MAE between 0.17 and 1.43. In our study (Table 3), $R^2$ ranged between 0.04 and 0.66 (lower maximum that Kaundal's), whereas MAE was between 0.16 and 1.20. However, Kaundal et al. [26] focused just on the agreement between the models signal and the incidence of the disease, whereas we also considered an 'early warning' metric, which is crucial in case of operational use of the models for providing farmers with timely warnings, thus allowing them to take effective remedial measures. Also, we have compared process models as well whereas Kaundal's paper only considered ML and statistical models. Taking into account the real difficulties of in-field data capture and how the blast severity itself is evaluated, we feel that our precision is realistic aligned with in-field predictability, and also taking into account the diverse geographical locations we have studied. The study of Kaundal was based on different locations (Palampur, Malan and Pharer) but which were very close (within 10kms of each other) in a pre-Himalaya region of North Western India. In our study, the locations are in Greece, Spain and Portugal, respectively, and therefore we would say that our project represents a much greater challenge in terms of different locations.

In our study, the best machine learning model (CNN) achieved an average $R^2$ of 0.50, and the best process model was WARM with an average $R^2$ value of 0.57 (Table 4). In terms of the "early warning" metric, the best methods were the ML models with average values of 181 and 112 for RNN and M5Rules, respectively, whereas the process models had average values of 88.29 and 9.29 for Yoshino and WARM, respectively.

However, we would also highlight that the in-situ capture of the rice blast severity index is clearly a critical factor which affects the data modelling, as this is the output variable used for supervised learning. The blast severity indicator for our work was captured as part of an EU FP7 project based on the adoption of state-of-the-art in-situ metrology equipment. It is clear that the comparison with other studies could be affected by the reliability of the methods used to evaluate rice blast incidence.

In the following we will now discuss the results from different viewpoints: (i) time of the crop season, (ii) meteorological differences between sites and (iii) dataset requirements.
**(i)** From Figs. 5 and 6 it can be seen that the ML models triggered early in the season (last weeks of June and first weeks of July) and then stayed above a certain level. On the other hand, the WARM and Yoshino models had a more "spiky" behavior although they continue to give a strong signal at specific points later in full summer (through to the end of August). This issue can lead us to ask questions in the light of the potential implications in terms of supporting farmers. For example, if the performances were poorer in a certain part of the season, is that part of the season particularly critical or maybe not in terms of potential impact of the pathogen? As a reply it could be stated that a poorer performance obtained later in the season should not penalize so much given the crop is less susceptible.

**(ii)** With respect to differences in performance between the sites, Table 6 shows the meteorological statistics for each site, and it can be see that Seville and Portugal had lower minimum temperatures and higher minimum relative humidity. Also, Seville had a lower max. temperature and higher leaf wetness. Relating this to the predictive results, from Tables 1 to 4 it can be seen that Portugal and Seville gave the relatively lowest model precisions as test datasets and this correlates with their relative variance in meteorological behaviour with respect to the Kalochori site.

**Table 6** Comparison of sites in terms of meteorological statistics.

|       | Temperature | | | | Relative humidity | | | | Leaf wetness | | | |
|---|---|---|---|---|---|---|---|---|---|---|---|---|
|       | Min. | Max. | Mean | StDev | Min. | Max. | Mean | StDev | Min. | Max. | Mean | StDev |
| **k2015** | 15   | 43   | 27.85 | 7.84 | 10   | 94   | 68.6  | 21.92 | 300 | 620 | 379.7 | 89.86 |
| **k2016** | 14.2 | 41.7 | 26.4  | 7.15 | 14.9 | 100  | 72.51 | 25.07 | 300 | 615 | 387.2 | 92.83 |
| **p2015** | 10   | 42   | 22.46 | 8.87 | 27   | 100  | 73.25 | 27.49 | 300 | 560 | 395   | 103.1 |
| **s2016** | 11.9 | 36.8 | 24.86 | 6.93 | 25.7 | 99.2 | 75.64 | 21.12 | 336 | 681 | 478   | 57.5  |

With regard to if process based models (in particular the Yoshino one that was developed in Japan, see description and parameter ranges in Methods section) worked better in one site than in another, from Table 4 it can be seen that the Yoshino model gave better performance in Kalochori than in Seville, and this may be related to the site being more similar (in terms of climate conditions) to the conditions experienced by the pathogen in Japan. This information could be used in order to re-parameterize the Yoshino model to better adapt it to the Portugal and Seville sites.

**(iii)** In terms of the size of the dataset (number and quality of observations) needed to develop machine learning approaches, the data was collected daily between May and September for each year and location, with a sampling frequency of 15 minute or 1 hour giving between 9600 and 2400 records in total. The key parameters captured in-situ by sensors were meteorological data (temperature, relative humidity and leaf wetness) as well as the "blast severity" (detected incidence of rice blast on the rice leaves, see Figure 1). This could give an idea of how many observations are necessary to develop similar models but under conditions different from the ones we have described in this paper. The quality of the observations was ocasionally impaired/reduced by data communication problems, and sensor failures, for example. More details of data capture and volumes are given in the "Methods- Data collection" section of the paper. Missing values (e.g. due to sensor failures) can be mitigated by interpolating existing values, when sufficient exist.

Future work could include trying to translate (at least in some cases to give an example) model errors in potential damages. That is, in the case that an "alert" (and thus the spraying treatment) would have been, for example,3-days late, how much more damage would have been suffered? This would relate errors at the time of spraying and damages to the crop (yield losses).

## Conclusions

For the first time, we have compared different process based models and machine learning approaches for their capability to support disease management. Given the specific objective, besides using standard agreement metrics such as R, $R^2$ and %MAE to evaluate the model reliability in simulating the incidence of rice blast, we defined and applied metrics (AUC, Area Under Curve) specifically targeting the evaluation of the suitability of the models in anticipating the appearance of the rice blast symptoms. In light of an operational use of the models, this is of fundamental importance to allow the timely application of countermeasures.

Among the process-based models, the Yoshino approach achieved performances that were slightly better than the WARM one, although WARM was fed using 2 km × 2 km gridded weather data, whereas Yoshino used the weather data collected using the in-field wireless sensor network developed within the RICE-GUARD project. This also underlines the importance of systems for the collection of in situ weather data for the simulation of diseases, given their higher representativeness.

The M5Rules machine learning approach used input data constructed from different moving averages (1, 3, and 7 days) to obtain triggers for detecting the conditions that anticipate the onset of Rice Blast disease. The RNN neural network learner provided a more intense signal that nevertheless coincided with the lead up period and the incidence period of the rice blast onset. Our results showed that all the approaches evaluated gave an early warning signal **before the appearance of symptoms**, thus making possible the adoption of effective preventive actions, in turns allowing to reduce crop losses and to minimize the use of fungicide spraying. In the case when high quality datasets with observations are available, machine learning approaches are much more flexible and easier to develop/parameterize. Otherwise, process based models could be the solution. Concerning the latter, the Yoshino approach demonstrated a great effectiveness in exploiting the availability of in situ, highly representative weather data, whereas WARM proved to be robust in case of less representative gridded weather data. However, both process models were susceptible to a lack of rainfall data. When leaf wetness was used as input as a substitute to rainfall, the process models failed to produce an output for the Kalochori 2015 and Portugal 2015 datasets.

Despite the practical difficulties, the results obtained are promising, and future studies will be carried out to further validate the approaches, e.g., by verifying the presence of false positives and testing the models in other rice production districts.

**Methods**
*Data collection*
The data was collected using a Wireless Sensor Network (WSN) system composed of three main elements:
1) the Master Node, responsible for collecting environmental measurements from the surrounding field environment and for receiving the data from the paddy nodes in the field which is then uploaded through a Wireless Area Network (WAN) to a remote database. It is formed by a) the Sensor and Supply Unit (SSU) which is the part of the master node that takes measurements from the different environmental sensors and the data is transmitted to b) the Control and Communications Unit (CCU), which uploads it to the remote database. The c) Network Communications Unit (NCU) is responsible for collecting data from paddy nodes through radiofrequency (RF) communications and sending it to the CCU. The master node contains sensors to measure temperature, relative humidity, barometric pressure, solar radiation, leaf wetness, rain and wind sensors.
2) the Paddy Node, responsible for taking environmental measures in the rice field and sending them to the master node through RF. Additionally, it can be used as a data logger that saves all data in an internal non-volatile memory to be downloaded through a Bluetooth connection using any Android device. This type of node can measure measure relative humidity, temperature and leaf wetness at four height levels and irradiance at three levels. The three bottom levels are intended to be within the rice canopy and the top level measured the external conditions on top of the rice level. These nodes are powered by an autonomous energy harvesting system composed of two solar panels which were capable of charging batteries as well as to measure diffuse irradiance. These nodes also acted as data loggers and/or repeater nodes which relay the information from other nodes in the field to the Master Node.
3) The cloud platform, where all the data was stored for later usage and access through a dedicated User Interface which allowed to see the location of the nodes, sensor readings, data trends and weather forecast.

The system described above was a custom system developed within the scope of the Rice-Guard project to provide a low-cost WSN made from off-the-shelf components with a potential commercial exploitation due to its performance and cost efficiency.

The installation of the system consisted of planting up to 3 Paddy Nodes in the each of the sampling locations together with other nodes acting only as repeaters to relay the information to the Master Node as well as 1 or more units of the commercial RH and Temperature sensor Hobo U21-001 for data validation purposes as seen in Figure 7.

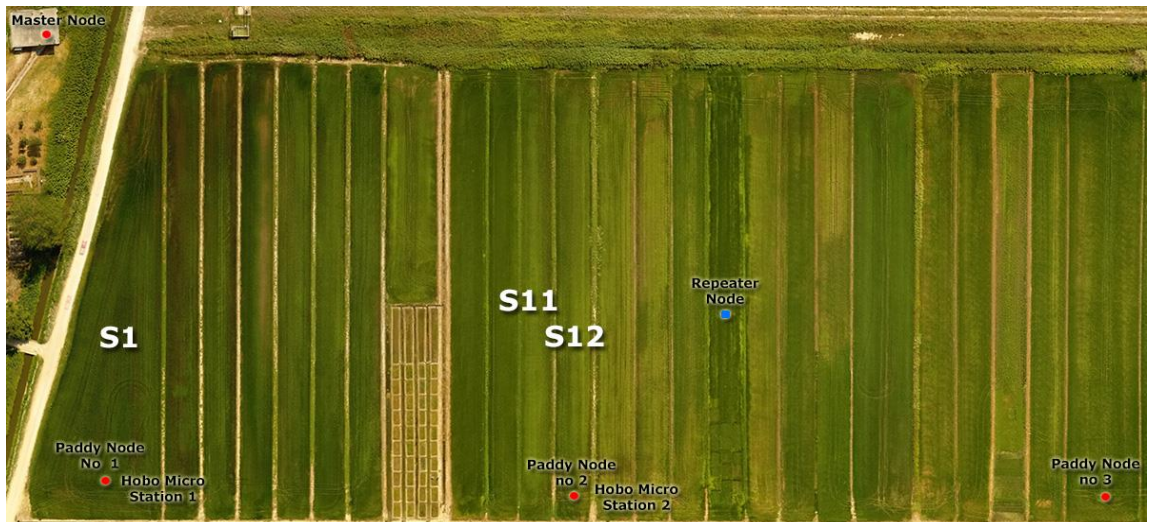

**Figure 7.** Location of the Master Nodes, Paddy Nodes and Hobo sensors in the experimental plots of Kalochori, Greece in 2016.

Thus, data was collected using the custom aforementioned system, together with commercial dataloggers Hobo U21-001. All the data was stored in the cloud either via manual downloading of the data from the dataloggers or nodes without connection to the WSN or automatically through network connectivity with the remote server.

Data validity was judged from the coherence between results from immediately close sensors throughout time, and their correlation with results from nearby master nodes and weather stations when available. A high resolution handheld meter was also used to validate values in the field. In these analyses, potential reading variability was accounted for. Similarly, outlier removal was performed when analysing the data sets after the field trials and prior to assessing rice-blast incidence based on data collection.

The data sampling areas were the located in paddy fields in Isla Mayor, Seville (Spain; 37°2' N, 6°6'W) between 15th June and 21st September 2016 and Kalochori (Greece; 40°36'N, 22°49'E) between 25th June and 14th September 2015/2016 (for the totality of the nodes). Also, the data from the Hobo commercial sensors was obtained from Montemor-o-Velho (Portugal; 40°08'N 8°38'W) between 5th of May and 20th of September 2015.

As for data volumes, from the Master Nodes, the data from Isla Mayor belongs to the aforementioned period with a sampling frequency of 1 measure every 15 minutes (9600 values). Data from Kalochori is also from the period described above with the same sampling frequency (7250 values). From the Paddy Nodes, the sampling frequency was 1 measurement every hour (cca. 2400 values).

### *Input data and pre-processing*
Four main datasets were used for the analysis, containing RICE-GUARD telemetry data at 10-minute temporal resolution and real leaf blast severity index. The four datasets were collected from paddy fields at three sites at Isla Mayor, Seville (s2016), Kalochori, Greece (k2015, k2016) and Portugal (p2015) between June and September 2015 and 2016. RICE-GUARD telemetry data referred to air temperature (°C), relative humidity (%), leaf wetness (-), and wind speed (m s$^{-1}$). All datasets were provided with relative humidity and temperature time series readings from which we calculated moving averages over one, three and seven days. Moving averages were used only for the M5Rules data modelling and not for the neural networks.

In the case of the machine learning algorithms, different combinations of the datasets were used for training/building the data models on the one hand, and data model evaluation on the other hand.

For the neural network, the sequence prediction problem was reframed as a supervised learning problem. That is, sequence series data were transformed from a sequence to pairs of inputs and outputs. Input at time *t* was a vector that contained values of input variables of *n* previous time steps, and the output was the flag at time *t*. The value of *n* (100) was defined by finding the best compromise between the system capability to capture the necessary time dependencies in a sequence and the containment of the training time. The scales of the different input variables were different, thus the input variables were normalized to the range [0,1]. This allowed to further speed up the learning process.

### *Rice blast prediction models*
Four approaches for rice blast prediction were evaluated: two process-based (Yoshino and WARM) and two based on machine learning techniques (M5Rules Rule Induction and RNN Neural Networks).

### *(i) Process-based models*
The Yoshino model [27, 14] was developed as a leaf blast forecasting model in Japan, and it is still in use in a variety of models or alerting systems. The model estimates the potential of hourly weather data to generate rice blast successful infections based on three rules: (i) mean air temperature of the past five days is between 20 and 25°C; (ii) rainfall intensity is lower than 4 mm h$^{-1}$; (iii) the continuous wet period is 4 hours more than the base wet hours, with the latter estimated from air temperature in wet hours. Once the three rules allow identifying infection hours, they are cumulated to calculate the daily infection warning hours (DIWH). DIWH is classified as: (i) no risk (DIWH = 0 h); (ii) low risk (1 h ≤ DIWH < 3 h); (iii) medium risk (3 h ≤ DIWH < 6 h), and (iv) high risk (DIWH ≥ 6 h).

The rice model WARM [28] includes a module for the simulation of damages due to leaf and panicle blast, successfully parameterized and tested in temperate regions [29] and currently used in Italy within a series of operational alert services (two requested by regional authorities and one by an insurance company). After the day of disease onset (estimated based on hydrothermal time), the daily infection efficiency is estimated according to Magarey et al. [34] as a function of hourly air temperature and leaf wetness duration. Weather variables needed for infection simulation are air temperature, relative humidity, leaf wetness, wind speed, and rainfall. WARM includes routines for the simulation of the whole disease progress, including reduction in green leaf area and translocation to grains, as well as in final yield. However, these processes were not considered within the current study.

Details on the Yoshino and WARM models, as well as on their parameterization and performances, are available in the reference literature.

### *(ii) Machine learning approaches*
M5Rules [35] is a tree induction algorithm which generates a decision list for regression problems using separate-and-conquer. In each iteration, it builds a model tree and makes the "best" leaf into a rule. M5Rules is an optimized algorithm for inducing simple, accurate decision lists from model trees. Model trees are built repeatedly, and the best rule is selected at each iteration. This method produces rule sets that are as accurate but smaller than the model tree constructed from the entire dataset. However, a trade-off is necessary between rule accuracy and rule coverage. Its reported performance makes it one of the best state of the art algorithms for rule induction where the output (predictive/classifier) variable is of numerical continues type. Figure 3 shows an example of an induced tree and associated rules. It is incorporated in the widely used "Weka" data mining software, made available from the University of Waikato, New Zealand.

A key part of a tree/rule induction algorithm is the "information gain measure" [36]. In the context of the partition of the training data set, the heuristic has a key dependence on an information gain calculation to evaluate which attribute to incorporate next, and where to incorporate it in the induction tree.

Let T be a set of training examples, each of the form $(x, y) = (x_1, x_2, x_3, ...., x_k, y)$ where $x_a \in vals(a)$ is the value of the $a^{th}$ attribute of example $x$ and $y$ is the corresponding class label. The information gain for an attribute $a$ is defined in terms of entropy H() as follows:

$$IG(T, a) = H(T) - \sum_{v \in vals(a)} \frac{|\{x \in T | x_a = v\}|}{|T|} \cdot H(\{x \in T | x_a = v\})$$

The mutual information is equal to the total entropy for an attribute if for each of the attribute values a unique classification can be made for the result attribute. In this case, the relative entropies subtracted from the total entropy are 0.

Neural networks, on the other hand, come in different types, such as standard NNs, convolutional NNs, recurrent NNs and different combination of these. The choice of which one to use depends on the specific application. We use Recurrent NNs [37,38] to model the appearance of the rice blast given the time sequences of weather parameters. RNNs are suitable for modeling time sequences, because they have loops where an output returns to an input (Figure 8.a) that allows them to "remember" the past.

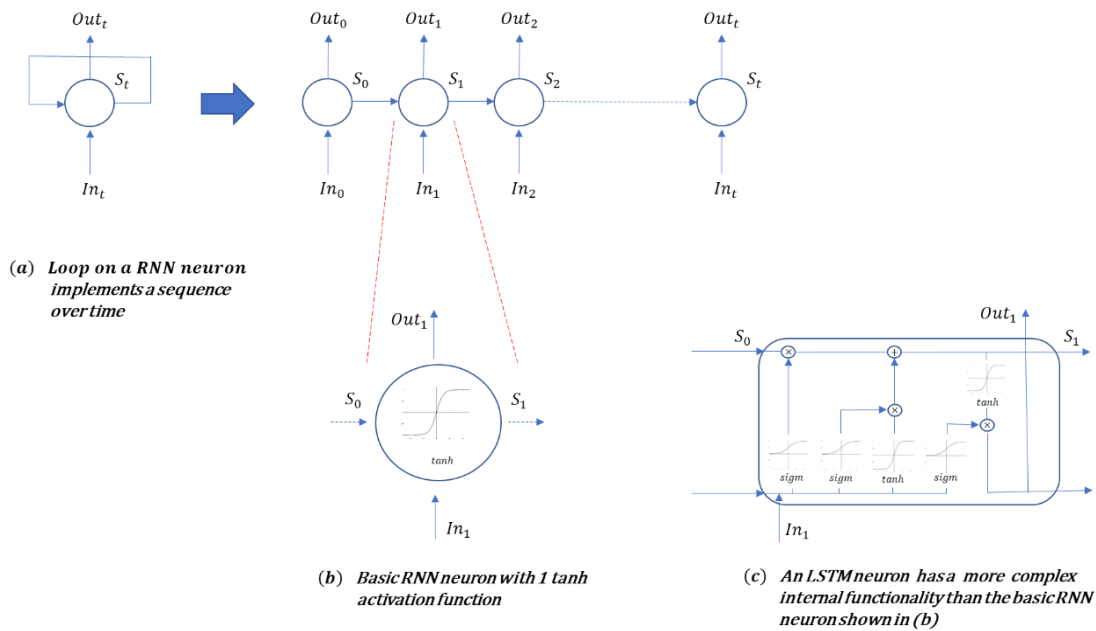

(a) Loop on a RNN neuron implements a sequence over time

(b) Basic RNN neuron with 1 tanh activation function

(c) An LSTM neuron has a more complex internal functionality than the basic RNN neuron shown in (b)

**Figure 8.** Recurrent Neural Networks (RNNS) vs Long Short Term Memory (LSTM) Neural Networks

More precisely, we used a special type of RNNs called Long Short Term Memory (LSTM) [39,40]. LSTMs have shown to be more effective than standard NNs and RNNs in many scenarios. This is because they can selectively remember patterns for long time windows. Figure 8.b shows one cell in a standard RNN network, while Figure 8.c shows one cell in a LSTM network. It is possible to notice that the LSTM cell is more complex than that of the standard RNN network. This specific structure of the LSTM cell allows it remembering and forgetting specific patterns through long time periods. This structure also avoids a vanishing gradient problem, thus allowing the training of deep LSTM networks.

The NNs are often referred to as "black box models". This means that, although they are good in capturing complex nonlinear relationship between input and output variables, it is difficult to interpret and understand their results, i.e., it is difficult to find human understandable rules of the conditions for the particular output. This is in contrast to the M5Rules model, and we considered useful to test both approaches in our analysis.

**Availability of data and materials**
The website of the official RICE-BLAST EU FP7 project website is available at: www.rice-guard.eu. In this website, we have provided a public link to the datasets used in this paper (one for training and one for test) together with an explanatory 'Readme' document, which is available at: http://multisite.iris.cat/riceguard/files/2018/09/DATASETS.zip.

**Competing Interests**
We declare that there are no competing interests and that all authors have approved the manuscript for submission.

**List of abbrevations used**
RNN, Recurrent neural networks; LSTM, Long short-term memory; WARM, Water Accounting Rice Model; MARS, Monitoring Agricultural ResourceS; PCA, Principal components analysis; DA, Discriminant analysis; MLRA, Multiple logistic regression analysis; MLP, Multilayer perceptron; ANN, Associative neural network; SVM Support vector machine; MSE, Mean squared error; MAE, Mean absolute error; WSN, Wireless sensor network; DIWH, Daily infection warning hours.

**Authors contributions**
DN carried out the M5Rules method data pre-processing, analysis, modelling and interpretation, as well as the overall writing, preparation and revision of the manuscript. DK and AK performed the in-field data capture, pre-processing and execution for the Yoshino Model, participating in the RICE-GUARD project, reviewing of the manuscript and contribution to the state of the art. NSD carried out the RNN method data pre-processing, analysis, modelling and interpretation, as well as devising the metrics for warning success. PP participated in the setup of the in-field telemetry equipment, data capture and pre-processing and project management of the RICE-GUARD project. RC performed the in-field data capture, processing and execution of the WARM Model, participating in the RICE-GUARD project, as well as fully revising the manuscript and making suggestions for improvement.

**Acknowledgements**
The RICE-GUARD project has received funding from the European Union's Seventh Framework Programme for research, technological development and demonstration: Under grant agreement nº 606583. 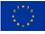

**Figure 1.** Rice Blast - different grades of leaf lesion (source: http://journals.plos.org/plosone/article?id=10.1371/journal.pone.0026260 )

**Figure 2.** Data capture RICE-GUARD station located outside the paddy field, for gathering and transmitting in real time readings from the in-field sensors

**Figure 3a.** Complete rule set of M5Rules model trained on the k2015+s2016+p2015 datasets

**Figure 3b.** Tree representation of rule model shown in Figure 3a

**Figure 4**. AUC (Area Under Curve) used as an "early warning" metric evaluator for the period 20th June to 7th July for the Kalochori 2016 dataset.

**Figure 5**. Real blast severity vs predicted: Train k2015+s2016+p2015, Test k2016

**Figure 6.** Real blast severity vs predicted: Train k2016+k2015+p2015, Test s2016

**Figure 7.** Location of the Master Nodes, Paddy Nodes and Hobo sensors in the experimental plots of Kalochori, Greece in 2016.

**Figure 8.** Recurrent Neural Networks (RNNS) vs Long Short Term Memory (LSTM) Neural Networks